\documentclass[slac_one]{revtex4}
\usepackage{graphicx}
\usepackage{fancyhdr}
\newcommand\sss{\scriptscriptstyle}
\newcommand\sQ{{\sss Q}}

\newcommand\as{\alpha_{\rm\scriptscriptstyle S}}

\def\asb{{}\ifmmode \bar{\alpha}_s \else $\bar{\alpha}_s$\fi}

\newcommand\mQ{m}
\newcommand\LambdaQCD{\Lambda_{\scriptscriptstyle \rm QCD}}

\def\beq{\begin{equation}}
\def\eeq{\end{equation}}

\def\beqn{\begin{eqnarray}}
\def\eeqn{\end{eqnarray}}

\def\({\left(}
\def\){\right)}

\def\Dnp{D_{\rm NP}}
\def\pDnp{\tilde{D}_{\pi}}
\def\gDnp{\tilde{D}_{\gamma}}

\def\mthr{m_{\rm thr}}

\begin{document}

\title{Power-suppressed effects in heavy quark fragmentation 
functions\footnote{Talk given by M. Cacciari at FRIF workshop on first principles 
non-perturbative QCD of hadron jets, LPTHE, Paris, France, 12-14 Jan 2006}}


\author{Matteo Cacciari}
\affiliation{LPTHE, Universit\'e P. et M. Curie - Paris 6, France}
\author{Paolo Nason}
\affiliation{INFN, Sezione di Milano, 
Piazza della Scienza 3, 20126 Milan, Italy}
\author{Carlo Oleari}
\affiliation{Universit\`a di Milano-Bicocca,
Piazza della Scienza 3, 20126 Milan, Italy}
%
%\author{P. Lucas}
%\affiliation{FNAL, Batavia, IL 60510, USA}

\begin{abstract}
This talk summarizes the results of a phenomenological analysis of heavy quark
fragmentation data published by the CLEO and BELLE
collaborations at $\sqrt{s} = 10.6$~GeV and by the LEP collaborations at
$\sqrt{s} = 91.2$~GeV. Several theoretical ingredients are employed: 
next-to-leading order initial conditions, evolution and coefficient
functions; soft-gluon resummation to next-to-leading-log
accuracy; a next-to-leading order matching condition for the crossing
of the bottom threshold in the  evolution. 
Important initial-state electromagnetic radiation effects in the
CLEO and BELLE data are also accounted for. We find that 
with reasonably simple choices of a non-perturbative
correction to the fixed-order initial condition for the evolution,
the data from CLEO and BELLE can be fitted with remarkable accuracy.
The fitted fragmentation function, when evolved to LEP energies,
does not however represent fairly the 
$D^*$ fragmentation spectrum measured by ALEPH.
Large non-perturbative corrections
to the coefficient functions of the meson spectrum are needed
in order to reconcile CLEO/BELLE and ALEPH results.

\end{abstract}

\maketitle

\thispagestyle{fancy}


\section{Introduction}

The CLEO~\cite{Artuso:2004pj} and the BELLE~\cite{Seuster:2005tr}
Collaborations have recently published high-statistics and high-accuracy data
for various charmed mesons fragmentation in $e^+e^-$ collisions at a
centre-of-mass energy of $10.6$~GeV. These data are the first ones able
to rival in quality (and, in fact, to best) similar ones published a few years
ago by the ALEPH Collaboration~\cite{Barate:1999bg} at $91.2$ GeV.

Taken together these sets of data allow for a phenomenological
analysis~\cite{Cacciari:2005uk} 
which spans a fairly large energy gap. It is therefore possible not only to
test the ability of the theoretical framework to describe the data well,
but also to perform the evolution from one energy to the other, and look for
evidence (or absence) of power suppressed effects.

The theoretical framework employed is based on a next-to-leading order 
QCD description of the fragmentation process~\cite{Mele:1990cw}, 
including a next-to-leading log accurate resummation of collinear 
and soft gluons~\cite{Cacciari:2001cw}\footnote{It is worth noting that very
recently next-to-next-to leading order 
results for the time-like non-singlet splitting function have become
available~\cite{Mitov:2006ic}. Together with the previously available ${\cal
O}(\alpha_s^2)$ initial conditions~\cite{Melnikov:2004bm} they make
possible to repeat at least part of this analysis with higher accuracy.}.

A few more details complete the theoretical picture. They are briefly
summarized below, and they are outlined in detail
in~\cite{Cacciari:2005uk}. 

First, and most important, the soft-gluon resummation needs to be
artificially regularized at large $x$ (or, equivalently, large $N$ in moment
space). The choice of the regularization defines the
perturbative distribution and therefore directly influences the non-perturbative
one extracted by fitting the experimental data. In this work we have decided to
use a prescription which, while being as simple as possible, possesses
the following desirable features:
\begin{itemize}
\item[(i)] it is consistent with all known perturbative results,
\item[(ii)] it yields physically acceptable results,
\item[(iii)] it does not introduce power corrections larger than
      generally expected for the processes in question, i.e.\ $N\Lambda/m$
      for the initial
      condition~\cite{Nason:1997pk,Jaffe:1993ie,Randall:1994gr,Cacciari:2002xb}
      and $N\Lambda^2/q^2$ for the coefficient
      functions~\cite{Dasgupta:1996ki}, where $\Lambda$ is a typical hadronic
      scale of a few hundreds MeV.
\end{itemize}
In practical terms, our choice will yield a fragmentation function which
does not become negative in the large-$X$ region. This will allow for a
good description of the data up to the the $x=1$ endpoint.

Second, while evolving the charm fragmentation function through the bottom
threshold, one needs in principle to modify the evolution equations and
properly match at this threshold. Moreover, production of charm via gluon
splitting should be allowed, since it represents a non entirely negligible
source at LEP energies. Both these features have been implemented in
\cite{Cacciari:2005uk}, thus departing from the simple non-singlet only description
previously usually employed. 

Finally, the experimental data as measured by CLEO and BELLE still contain the
effect of electromagnetic initial state radiation. We have estimated that this
effect is not negligible in this case (as it is, instead, at LEP, due to the
physical cutoff provided by the $Z^0$ resonance peak). We have therefore proceeded
to simulate it and to deconvolute it from the data before fitting them with a
pure QCD description of the fragmentation process. Again, details can be found
in \cite{Cacciari:2005uk}.

\section{Non-perturbative fragmentation function}\label{sec:nonpFF}
In the heavy-quark fragmentation-function formalism, the largest
non-perturbative effects come  
from the initial condition, since one expects power corrections
of the form $\Lambda/\mQ$.
We assume that all these
effects can be described by a non-perturbative fragmentation function
$\Dnp$, that takes into account all low-energy effects, including
the process of the heavy quark turning into a
heavy-flavoured hadron, that has to be convoluted with the perturbative cross
section.
Thus, the Mellin transform of the
full resummed cross section, including
non-perturbative corrections, is
\beq
\label{eq:hadfactor}
   \sigma_{\sss H}(N,q^2) = \sigma_\sQ(N,q^2,\mQ^2)  \Dnp (N) \;.
\eeq
We have attempted to fit CLEO and BELLE $D^*$ data using several forms for
$\Dnp$.
We found that the best fits are obtained with the two-component form
\begin{equation}
\Dnp(x)={\rm Norm.} \times \frac{1}{1+c}
\left[ \delta(1-x) + c N_{a,b}^{-1} (1-x)^a x^b\right]\;,
\label{eq:threepar}
\end{equation}
with
\begin{equation}
N_{a,b}=\int_0^1 (1-x)^a x^b\;.
\end{equation}
This form is a superposition of a maximally hard component (i.e.\ the delta function)
and the form proposed in Ref.~\cite{Colangelo:1992kh}.
It can be given a simple phenomenological interpretation,
the hard term corresponding in some sense to the direct exclusive
production of the $D^*$, and the Colangelo-Nason form accounting for
$D^*$'s produced in the decay chain of higher resonances.

Following
the approach  of Ref.~\cite{Cacciari:2003zu}, we assume that the $D$ meson 
non-perturbative fragmentation function is the sum of a direct component, which
is isospin invariant, plus the component arising from the $D^*$ decay.
The decay $D^*\to D\pi$ is very close to threshold, so that the $D$ has the same
velocity of the $D^*$, and their momenta are thus proportional to their masses.
Under these circumstances, the component of the $D$ fragmentation function
arising from $D^* \to D\pi$ decays is given by
\begin{equation}
B(D^*\to D\pi)\; \pDnp^{D}(x)\;,
\end{equation}
where we have defined
\begin{equation}
\pDnp^{D}(x) =  \Dnp^{D^{*}}\(x \frac{m_{D^*}}{m_{D}}\)\;\frac{m_{D^*}}{m_{D}}
\;\theta\(1-x \frac{m_{D^*}}{m_{D}}\)\,,
\end{equation}
and $B(D^*\to D\pi)$ is the branching ratio of $D^* \to D\pi$.
Observe that $\pDnp^{D}$ has been defined so as to have the same normalization
as $\Dnp^{D^{*}}$. In $N$ space we obtain immediately
\begin{equation}
\pDnp^{D}(N) =  \Dnp^{D^{*}}(N) \left[\frac{m_D}{m_{D^*}}\right]^{N-1}\;.
\end{equation}

For the $D^*\to D\gamma$ decay, in the $D^*$ frame,
the $D$ has non-negligible velocity,
but it is non-relativistic, its momentum being given by
\begin{equation}
p_D=\frac{m_{D^*}^2-m_D^2}{2m_{D^*}}\;.
\end{equation}
It can easily be shown~\cite{Cacciari:2005uk} that, in moment
space, we can write
\begin{equation}
\gDnp^{D}(N) 
= \Dnp^{D^*}(N)\, \frac{m_{D^*}}{2p_D}\, \frac{(m_D+p_D)^N-(m_D-p_D)^N}{N m_{D^*}^N}\;.
\end{equation} 
We thus describe $D^{+/0}$ production as the sum of a primary (i.e.\ not coming
from $D^*$ decays) component, plus the contributions coming from
$D^*$ decays
\begin{eqnarray}
\Dnp^{D^{+}}(x) &=& \Dnp^{D^+,p}(x)
 + B(D^{*+}\to D^{+}\pi^0) \pDnp^{D^{+}}(x)
\nonumber \\&&
 + B(D^{*+}\to D^{+}\gamma) \gDnp^{D^{+}}D(x)\;,
\\
\Dnp^{D^0}(x) &=& \Dnp^{D^0,p}(x)
 + [B(D^{*+}\to D^0\pi^+)+ B(D^{*0}\to D^{0}\pi^0)] \pDnp^{D^{0}}(x)
\nonumber \\ &&
+B(D^{*0}\to D^{0}\gamma) \gDnp^{D^{0}}(x)\;.
\end{eqnarray}

\section{\boldmath{$D$} mesons data fits near the \boldmath{$\Upsilon(4S)$}}
\label{sec:cleobelle}
Several parameters enter our calculations. First of all, at all matching
points, there are scale choices that could be varied, to yield a perturbative
uncertainty in our result. Those are the initial evolution scale $\mu_0$, the
matching scale for the crossing of the $b$ threshold $\mu_{\rm thr}$, and the
final evolution scale $\mu$.  In the present work we fix
\begin{equation}
\mu_0=m\,,\quad \mu=\sqrt{q^2}\equiv \sqrt{s}\,,\quad \mu_{\rm thr}=\mthr=m_b\;.
\end{equation}
These scales could, in principle, be varied by a factor of order two
around the values listed above, yielding a sensibly different
result. However, in general, the scale variation
will simply result in different values for the fitted parameters
of the non-perturbative form. When computing cross sections
for different processes, one should then use the parametrization
appropriate for the scale choice that has been made in the fit, hence
compensating for the change.
In the present work we will not pursue this issue further,
since our aim is simply to show that a fit within QCD is possible.
A similar remark applies to the value of $\LambdaQCD$ and the quark masses,
that we will fix at
\begin{equation}
\Lambda^{(5)}_{\rm QCD}=0.226\;\mbox{GeV}\,,\quad m_c=1.5 \;\mbox{GeV}\,,
\quad m_b=4.75\;\mbox{GeV}\,.
\end{equation}

\begin{table}[t]
\begin{center}
\begin{tabular}{|l|c|c|c|c|c|}
\hline
\multicolumn{6}{|c|} 
{Eq.~(\protect{\ref{eq:threepar}}): $a=1.8 \pm 0.2$, \ $b=11.3 \pm 0.6$, \ 
  $c=2.46\pm 0.07$, \ total  $\chi^2= 139 $ } 
\\ \hline
 Set &
(C) $D^{*+}$ & (B) $D^{*+}\to D^0$ & (B) $D^{*+}\to D^+$ &
(C) $D^{*0}$ & (B)  $D^{*0}$
\\ \hline
Norm. & $0.238$ & $0.253$ & $0.227$ &$0.225$ & $0.211$
\\ \hline 
$\chi^2/$pts & 33/16 & 63/46 & 13/46& 13/16 & 17/46
\\ \hline
\end{tabular}
\caption{\label{tab:fitdstar}
Results of the fit to $D^*$ CLEO (C) and BELLE (B) data.
The last line reports the $\chi^2$ over the number of fitted points
for each data set.}
\end{center}
\end{table}

\begin{figure}[htb]
\begin{center}
  \includegraphics[width=\textwidth]{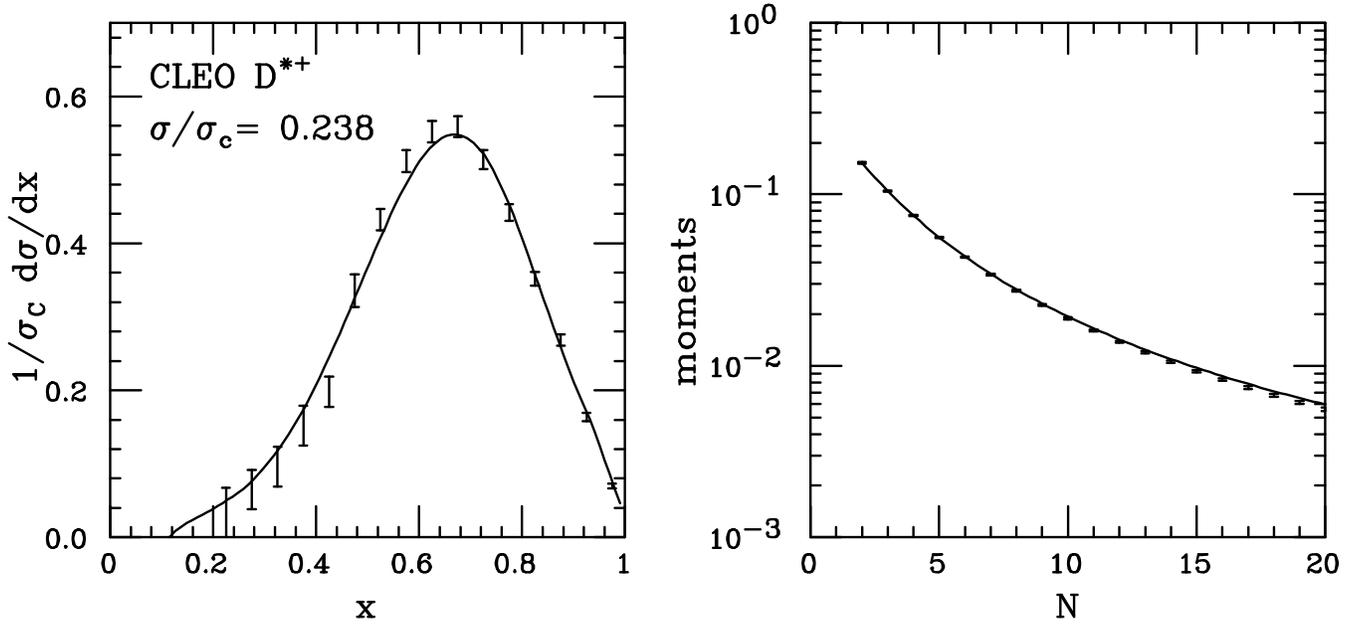}
\caption{\label{fig:CLEOdstarp}
Fit to CLEO $D^{*+}$ data.}
\end{center}
\end{figure}

\begin{figure}[htb]
\begin{center}
  \includegraphics[width=\textwidth]{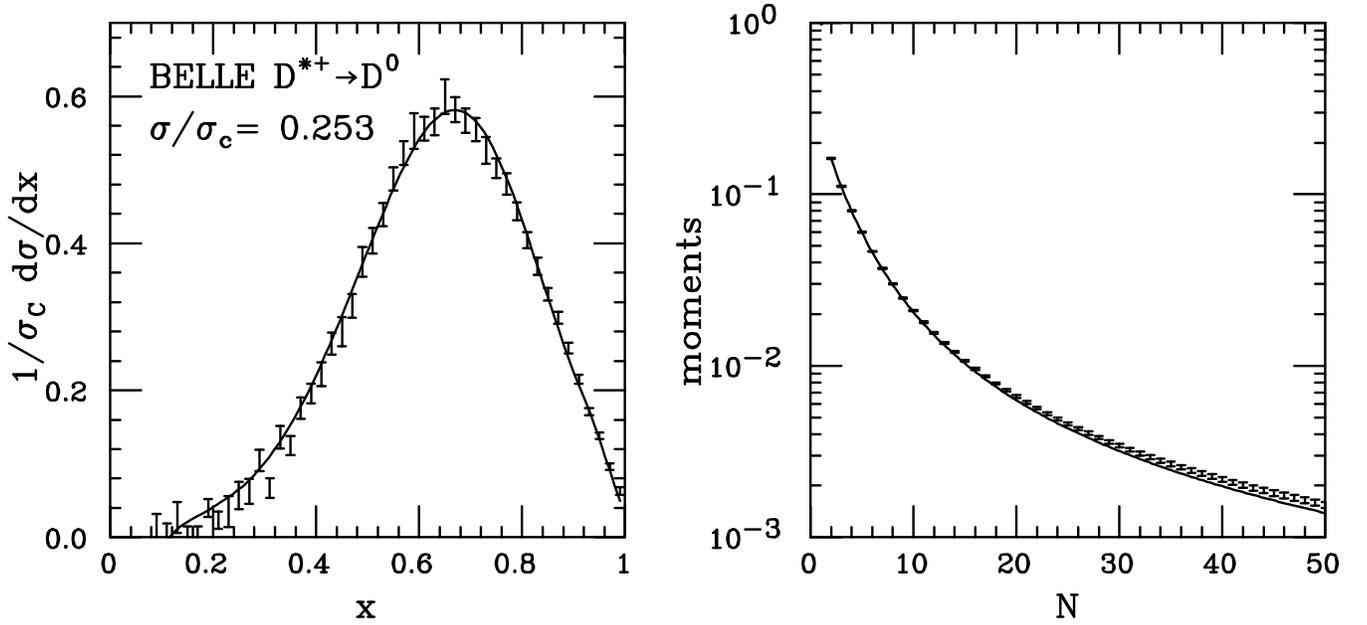}
\caption{\label{fig:BELLEdstarptoD0}
Fit to BELLE $D^{*+}\to D^0$ data.}
\end{center}
\end{figure}

The result of the fit is reported in Table~\ref{tab:fitdstar}, and
in Figs.~\ref{fig:CLEOdstarp}, \ref{fig:BELLEdstarptoD0}
%,\ref{fig:BELLEdstarptoDp}, \ref{fig:CLEOdstar0} and~\ref{fig:BELLEdstar0}
we show some of the data and the corresponding fitted curve, 
both in $x$ and moment space.
% \begin{figure}[htb]
% \begin{center}
%   \includegraphics[width=\textwidth]{BELLE-Dstarp-to-Dp.eps}
% \caption{\label{fig:BELLEdstarptoDp}
% Fit to BELLE $D^{*+}\to D^+$ data.}
% \end{center}
% \end{figure}
% \begin{figure}[htb]
% \begin{center}
%   \includegraphics[width=\textwidth]{CLEO-Dstar0.eps}
% \caption{\label{fig:CLEOdstar0}
% Fit to CLEO $D^{*0}$ data.}
% \end{center}
% \end{figure}
% \begin{figure}[htb]
% \begin{center}
%   \includegraphics[width=\textwidth]{BELLE-Dstar0.eps}
% \caption{\label{fig:BELLEdstar0}
% Fit to BELLE $D^{*0}$ data.}
% \end{center}
% \end{figure}

\section{\boldmath{$D$} mesons data fits on the \boldmath{$Z^0$}}
\label{sec:aleph}

\begin{figure}[htb]
\begin{center}
  \includegraphics[width=\textwidth]{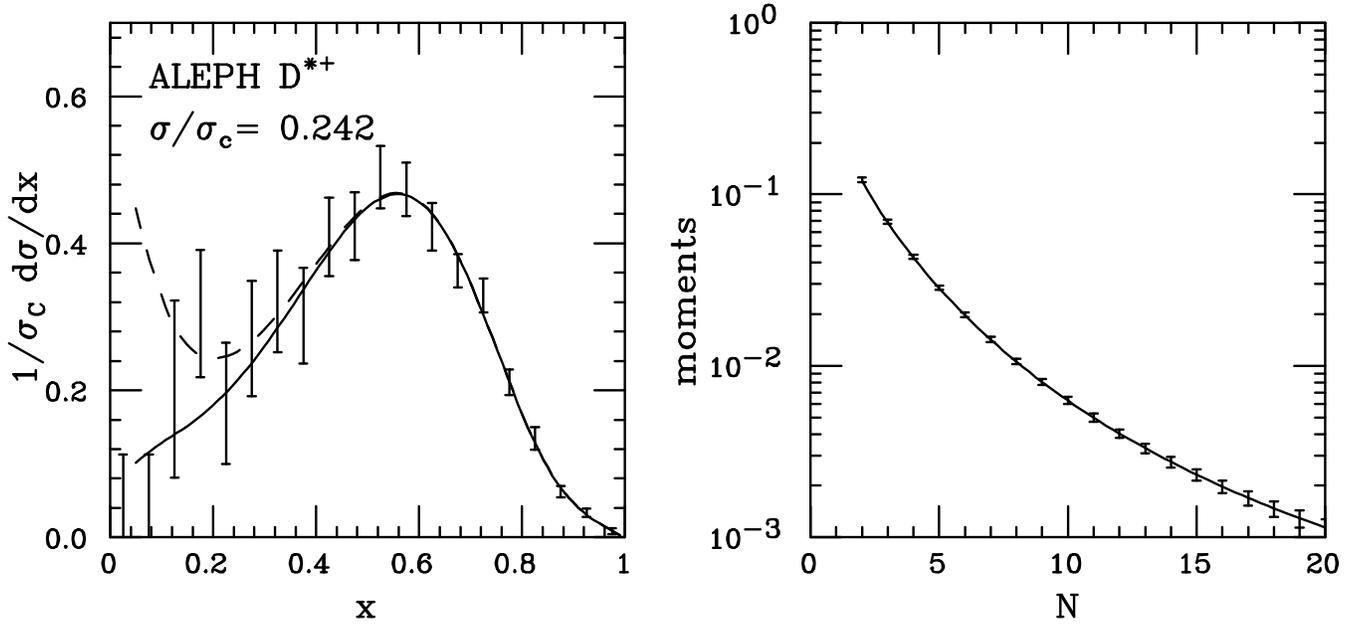}
\caption{\label{fig:ALEPH-Dstarp-fit}
ALEPH $D^{*+}$ data and the result of our non-singlet fit (solid line).
The dashed line represents the result obtained using full evolution.}
\end{center}
\end{figure}

Fig.~\ref{fig:ALEPH-Dstarp-fit} shows a similar fit to ALEPH $D^{*+}$ 
data~\cite{Barate:1999bg}.
We fit the data in the region $x \in [0.4,1]$ using the non-singlet component
only, since a subtraction of the gluon-splitting contributions was performed
by ALEPH.  Observe that, in this calculation, the bottom-threshold crossing
has to be dealt with.  We also show, for comparison, the full
evolution result (dashed line), using the same parameters obtained in the
non-singlet fit.  As we can see, the difference is only visible at small $x$.
The result of the fit for the non-perturbative parameters is
\begin{equation}\label{eq:ALEPHDstarfit}
a=2.4\pm 1.2\,,\quad  b=13.9\pm 5.7\,\quad c=5.9\pm 1.7\,,
\end{equation}
with a $\chi^2=4.2$ for 13 fitted points. These results are
not really consistent with
those for the $\Upsilon(4S)$ data in Tab.~\ref{tab:fitdstar}.

\begin{figure}[t]
\begin{center}
  \includegraphics[width=\textwidth]{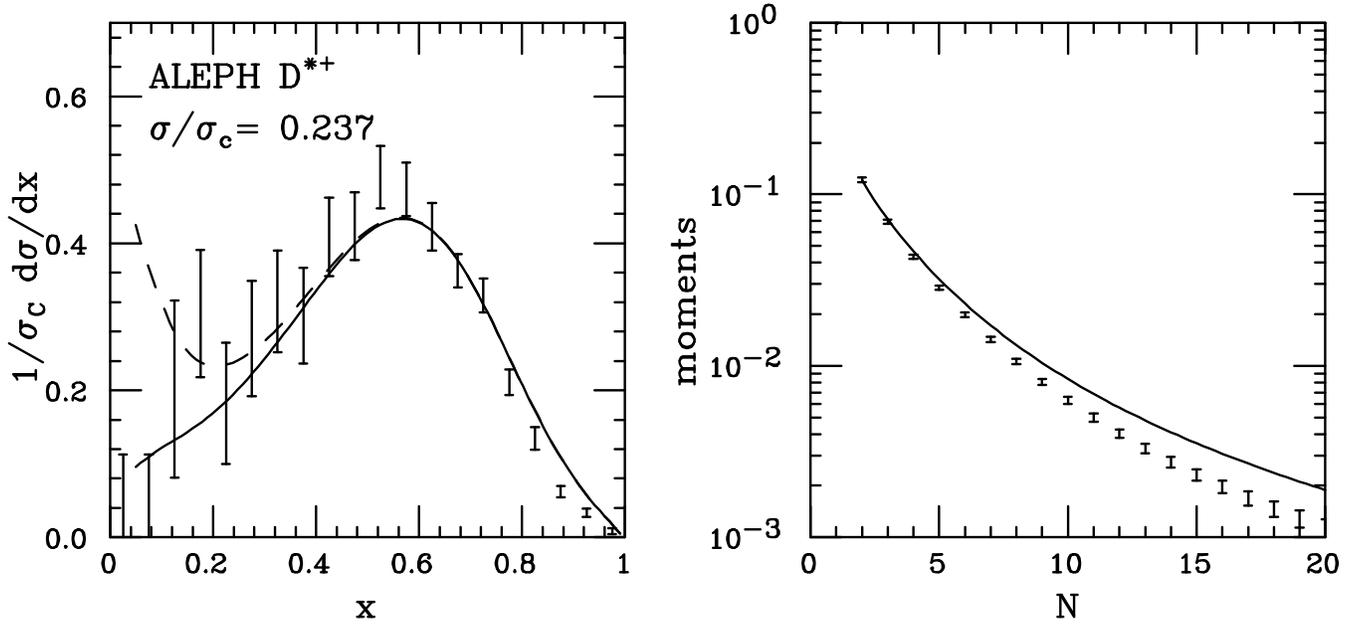}
\caption{\label{fig:ALEPHDstarp}
ALEPH $D^{*+}$ data, compared to the QCD prediction.}
\end{center}
\end{figure}

In order to better quantify the discrepancy between
Eq.~(\ref{eq:ALEPHDstarfit}) and Tab.~\ref{tab:fitdstar}
we use the parametrization of CLEO and BELLE data
to predict the $D^{*}$ fragmentation
function at LEP energies.
The LEP prediction, using the parametrization
of Table~\ref{tab:fitdstar}, is reported in
Fig.~\ref{fig:ALEPHDstarp} together with the ALEPH data.
We find a $\chi^2=60.1$ (for 13 fitted points) for this parametrization.
Thus, the description is not satisfactory, especially in the
large-$x$ (large-$N$) region.

\begin{figure}[ht]
\begin{center}
  \includegraphics[width=0.75\textwidth]{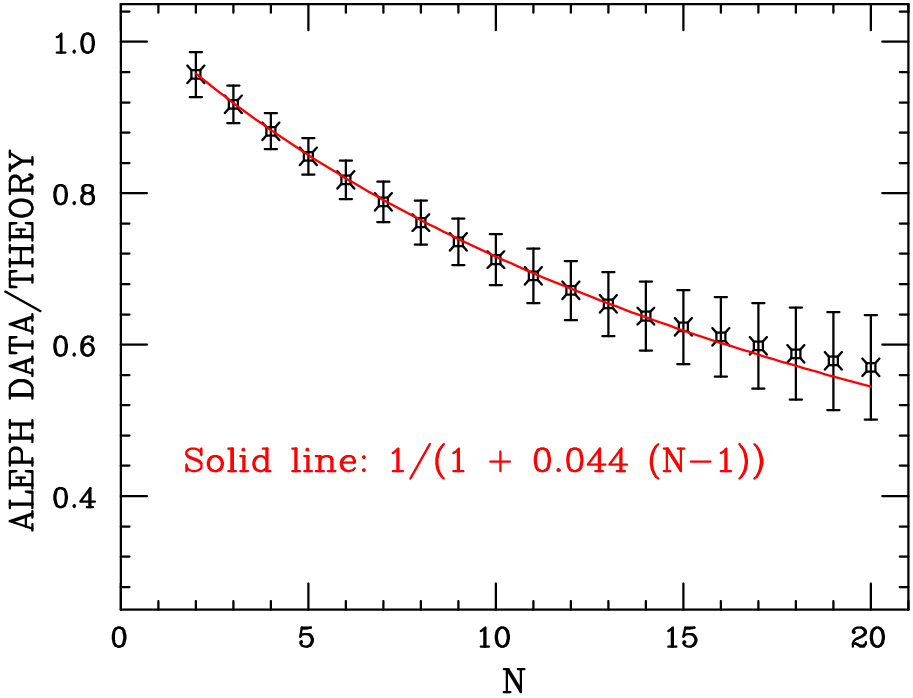}
\caption{\label{fig:ALEPHoTH}
ALEPH $D^{*+}$ data, compared to the QCD prediction.}
\end{center}
\end{figure}

In Fig.~\ref{fig:ALEPHoTH} we show the ratio of the moments
of ALEPH $D^{*+}$ data over our prediction.
We observe that the $N$ dependence of the ratio is well described
by the functional form
\begin{equation}\label{eq:depn}
\frac{1}{1+ 0.044 \,(N-1)}\;,
\end{equation}
where,
since the first
moment of the non-singlet distribution should be exactly
given by the theory (because of charge conservation),
we normalize to one the extrapolation of the data to $N=1$.

We can only speculate about the possible origin of the discrepancy and the
form of the coefficient of $(N-1)$ in Eq.~(\ref{eq:depn}).
Assuming that we are dealing with a non-perturbative correction
to the coefficient function of the form
\begin{equation}\label{eq:cfpc2}
1+\frac{C(N-1)}{q^2}\;,
\end{equation}
this would lead to the extra factor
\begin{equation}
      \frac{1+\frac{C(N-1)}{M_Z^2}}{1+\frac{C(N-1)}{M_\Upsilon^2}}\;,
\end{equation}
(where  $M_Z$ and $M_\Upsilon$ are the $Z^0$ and $\Upsilon(4S)$ mass)
to be applied to our prediction for the ALEPH data.
For $C=5~{\rm GeV}^2$ we reproduce the behaviour of
Eq.~(\ref{eq:depn}).
In Ref.~\cite{Dasgupta:1996ki}, on the basis of a calculation
of infra-red renormalon effects, a $1/q^2$ power correction is found,
with an $N$ dependence marginally compatible with~(\ref{eq:cfpc2}).
No $1/\sqrt{q^2}$ correction is found.
Ref.~\cite{Beneke:1997sr} also predicts a leading $1/q^2$ power correction.
 However, the
$C \approx 5~{\rm GeV}^2$ coefficient would appear to be somewhat too
large\footnote{If we believe that it is the maximum meson energy, not
$\sqrt{q^2}$,
that controls power effects, than we would have $C\approx 1~{\rm GeV}^2$,
a more acceptable value.}.
Alternatively, if we admitted the existence of corrections to the coefficient
functions of the form
\begin{equation}\label{eq:cfpc1}
1+\frac{C(N-1)}{\sqrt{q^2}}\;.
\end{equation}
then we would find $C\approx 0.52~{\rm GeV}$, a much more acceptable value.
We observe that a form
\begin{equation}
\left(1+\frac{C}{\sqrt{q^2}}\right)^{N-2} \approx 1 +
\frac{C(N-2)}{\sqrt{q^2}}
\end{equation}
was required
in Ref.~\cite{Nason:1994xx} to fit light-hadron fragmentation data.

Demonstrating the absence (or the existence) of $1/\sqrt{q^2}$ corrections
in fragmentation functions would be a very interesting result, since
it would help to validate or disprove renormalon-based predictions.
Unfortunately, the low precision of the available data  does not allow,
for the time being, to resolve this issue.

\begin{figure}[t]
\begin{center}
  \includegraphics[width=0.75\textwidth]{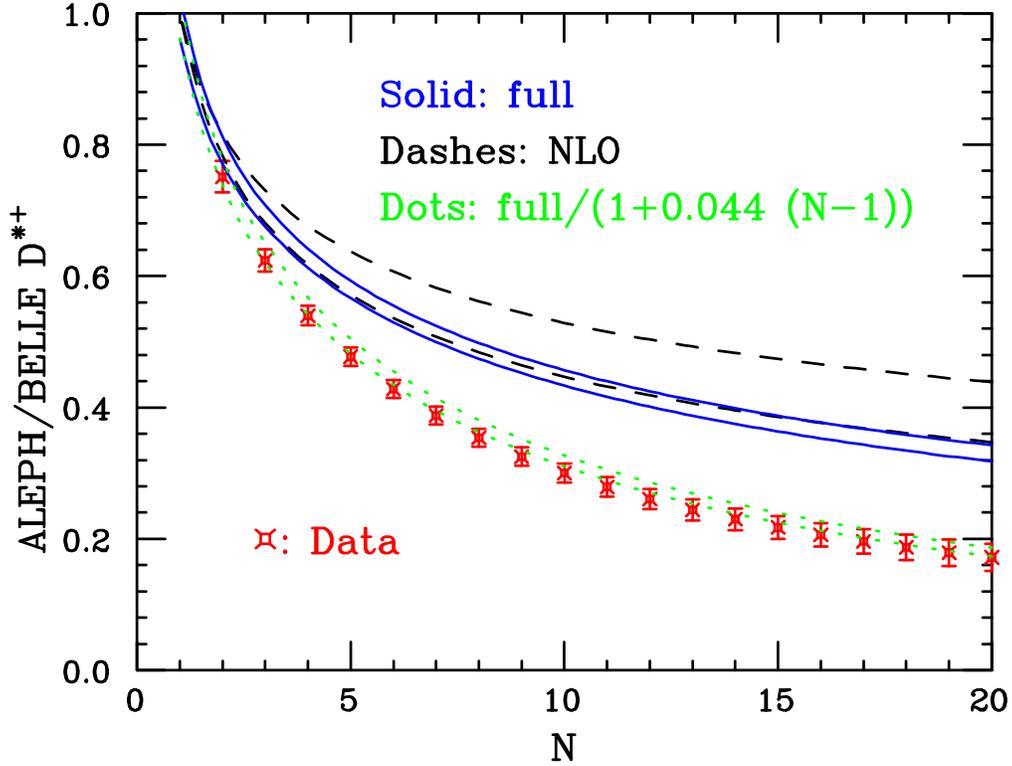}
\caption{\label{fig:ALEPHoBELLE}
The ratio of ALEPH and BELLE moments for the $D^{*+}$ fragmentation
function, compared to QCD evolution.  The solid band is obtained with
QCD NLO evolution and Sudakov effects in the coefficient functions,
while the dashed bands is NLO evolution only.
The bands are obtained by setting $\mu_{Z/\Upsilon}=\xi M_{Z/\Upsilon}$
and varying $1/2<\xi<2$.}
\end{center}
\end{figure}

We would like to remark that the discrepancy between the CLEO/BELLE and
ALEPH data exclusively depends upon the evolution between the $\Upsilon(4S)$
and $Z^0$ energies. The method we used to describe the CLEO/BELLE data
(i.e.\ the perturbative calculation of the fragmentation function,
the Sudakov effects in the initial conditions and the parametrization
of the non-perturbative part) does not affect the conclusions of the
present section.
 In fact, we can simply compute the ratio of the moments of the
inclusive $D^{*+}$ (ISR corrected) distribution at CLEO/BELLE and ALEPH,
and compare it to the theoretical prediction.
The result of this comparison (where we have used, for simplicity,
BELLE data only) is displayed in Fig.~\ref{fig:ALEPHoBELLE}.
The curves are given by
\begin{equation}\label{eq:evfac}
\frac{\sigma_\sQ(N,M_Z^2,m^2)}{\sigma_\sQ(N,M_\Upsilon^2,m^2)} = 
\frac{\bar{a}_q(N,M_Z^2,\mu_Z^2)}{
1+\as(\mu_{Z}^2)/\pi}
\;
E(N,\mu_Z^2,\mu_{\Upsilon}^2) \;
\frac{1+\as(\mu_{\Upsilon}^2)/\pi}{
\bar{a}_q(N,M_\Upsilon^2,\mu_{\Upsilon}^2)}
\end{equation}
where $\mu_Z$ and $\mu_\Upsilon$ are the factorization scales and
the evolution factor $E$ is given by the solution of the
Altarelli-Parisi evolution equation.
Notice that low-scale effects, both at the heavy
quark mass scale and at the non-perturbative level, cancel completely in
this ratio, making its prediction entirely perturbative.
% For $\bar{a}_q$, in the NLO results (dashed lines), we have used
% \begin{equation}
% \bar{a}_q(N,q^2,\mu^2)=1+\asb(\mu^2)\, a_q^{(1)}(N,q^2,\mu^2)\;,
% \end{equation}
% while for the full result (solid lines) we have included
% the NLL resummation of soft gluon emission in the coefficient functions
% \beqn
% \label{eq:coeffun_sud_noexp}
% \bar{a}_q(N,q^2,\mu^2)&=&\Delta_q^S(N,q^2,\mu^2) \nonumber\\
% &&\times \left\{  1+\asb(\mu^2) \left[a_q^{(1)}(N,q^2,\mu^2)
% -\lq\Delta_q^S(N,q^2,\mu^2)\rq_{\as}\right]\right\}. \phantom{aaa}
% \eeqn
% The definitions of $a_q^{(1)}$ and $\Delta_q^S$ are
% given in Sections~\ref{sec:Collinear_logarithms}
% and~\ref{sec:Soft_logarithms} of Ref.~\cite{Cacciari:2005uk}.
% We have set $\mu_{Z/\Upsilon}=\xi M_{Z/\Upsilon}$
% with $\xi=0.5,2$ to plot our bands. 
As we can see from the figure, the rather large scale uncertainty
displayed by the NLO result is much reduced when Sudakov effects
are included. In both cases, however, the data clearly undershoot
the pure QCD prediction, being instead compatible with the inclusion
of the correction factor~(\ref{eq:depn}) (dotted lines).
% We have also checked that our full result is essentially unchanged
% if, instead of formula~(\ref{eq:coeffun_sud_noexp}), we use
% the fully exponentiated formula~(\ref{eq:a_Q^res}).

One can legitimately wonder whether some of the theoretical ingredients
might hide further uncertainties.
We have checked that the regularization procedure needed
to deal with the Landau pole in the soft-gluon resummation 
has very little impact on our curves.
Using the very large value $\LambdaQCD^{(5)}=0.3\,$GeV would lower the
theoretical predictions by no more than 11\%{} for $N\leq 20$,
very far from explaining the observed effect.
The deconvolution of ISR effects, that hardens the $\Upsilon(4S)$ data, but is
insignificant on the $Z^0$, widens the discrepancy, which would however
still be partially visible even if the data were not corrected for e.m.
radiation.

Because of the relatively low energy of the data on the $\Upsilon(4S)$,
it is also legitimate to wonder whether charm-mass effects could be responsible
for the discrepancy between LEP and $\Upsilon(4S)$ data.
We have not included mass effects in the present calculation.
However, in Ref.~\cite{Nason:1999zj}, mass effects in charm
production on the $\Upsilon(4S)$ where computed at order $\as^2$,
and found to be small. In fact, they amount to a correction of the order
of  1{\%} at $N=5$, 4{\%} at $N=11$ and  7{\%} at $N=20$,
very far from being able to explain our observation.
We thus believe that it is
unlikely that mass effects 
could play an important role in explaining this discrepancy~\footnote{In
Refs.~\cite{Jaffe:1993ie} and~\cite{Nason:1997pk}, on the basis of the 
analogy with the spacelike case, corrections of the form $\Lambda m/q^2$ are
introduced. The importance of these corrections in the present framework
would require further investigation.}.

\section{Conclusions}
This phenomenological analysis~\cite{Cacciari:2005uk} of heavy quark 
fragmentation in $e^+e^-$
collisions shows that it is possible to perform excellent fits of $D^*/D$ meson
fragmentation spectra in perturbative QCD, using all known results on the
perturbative heavy-quark fragmentation function, and compounding them with a
simple parametrization of non-perturbative effects. 

A second striking result is the evidence of large non-perturbative effects,
visible in the relation between the $D^*$ fragmentation function at the
$\Upsilon(4S)$ and $Z^0$ energies.  It would be interesting to understand the
power law of these contributions. Their magnitude would suggest a
$1/\sqrt{q^2}$
scaling law. Theoretical arguments based upon infrared renormalons would
favour, instead, a $1/q^2$ behaviour. Because of the lack of precise
$D^*$ production data in the intermediate region, it is difficult, at 
this point,
to discriminate between the two possibilities.  We point out, however, that,
if these non-perturbative corrections involve the coefficient functions, they
may be present also in light-hadron production, where data at intermediate
energy are available.  It is thus possible that fits to the light-hadron
fragmentation functions from $\Upsilon(4S)$ up to $Z^0$ energies may clarify
this issue.
\label{sec:Conc}

\bibliography{paper}

\end{document}